\documentclass[sigconf]{acmart}
\usepackage{algorithm}
\usepackage{algorithmic}
\usepackage{cleveref}
\usepackage{balance} 
\AtBeginDocument{%
  \providecommand\BibTeX{{%
    \normalfont B\kern-0.5em{\scshape i\kern-0.25em b}\kern-0.8em\TeX}}}

\copyrightyear{2024}
\acmYear{2024}
\setcopyright{acmlicensed}
\acmConference[KDD '24] {Proceedings of the 30th ACM SIGKDD Conference on Knowledge Discovery and Data Mining }{August 25--29, 2024}{Barcelona, Spain.}
\acmBooktitle{Proceedings of the 30th ACM SIGKDD Conference on Knowledge Discovery and Data Mining (KDD '24), August 25--29, 2024, Barcelona, Spain}
\acmISBN{979-8-4007-0490-1/24/08}
\acmDOI{10.1145/3637528.3671645}

\begin{document}

\title{Non-autoregressive Generative Models for Reranking Recommendation}

\author{Yuxin Ren}
\affiliation{%
  \institution{Kuaishou Technology}
  \state{Beijing}
  \country{China}
}
\email{renyuxin@kuaishou.com}

\author{Qiya Yang}
\affiliation{%
  \institution{Peking University}
  \city{Beijing}
  \country{China}}
\email{yangqiya@stu.pku.edu.cn}

\author{Yichun Wu}
\affiliation{%
  \institution{Tsinghua University}
  \city{Beijing}
  \country{China}}
\email{wuyc21@mails.tsinghua.edu.cn}

\author{Wei Xu}
\affiliation{%
  \institution{Kuaishou Technology}
  \city{Beijing}
  \country{China}}
\email{xuwei09@kuaishou.com}

% \author{Yunhao Li}
% \affiliation{%
%   \institution{Kuaishou Technology}
%   \city{Beijing}
% \email{liyunhao@kuaishou.com}

\author{Yalong Wang}
\affiliation{%
  \institution{Kuaishou Technology}
  \city{Beijing}
  \country{China}
}
\email{wangyalong03@kuaishou.com}

\author{Zhiqiang Zhang$^\dagger$}
\thanks{$\dagger$ \text{Corresponding author}}
\affiliation{%
 \institution{Kuaishou Technology}
 \city{Beijing}
 \country{China}}
\email{zhangzhiqiang06@kuaishou.com}
\renewcommand{\shortauthors}{Yuxin Ren et al.}

\begin{abstract}
 Contemporary recommendation systems are designed to meet users' needs by delivering tailored lists of items that align with their specific demands or interests. In a multi-stage recommendation system, reranking plays a crucial role by modeling the intra-list correlations among items. The key challenge of reranking lies in the exploration of optimal sequences within the combinatorial space of permutations. Recent research proposes a generator-evaluator learning paradigm, where the generator generates multiple feasible sequences and the evaluator picks out the best sequence based on the estimated listwise score. The generator is of vital importance, and generative models are well-suited for the generator function. Current generative models employ an autoregressive strategy for sequence generation. However, deploying autoregressive models in real-time industrial systems is challenging. Firstly, the generator can only generate the target items one by one and hence suffers from slow inference. Secondly, the discrepancy between training and inference brings an error accumulation. Lastly, the left-to-right generation overlooks information from succeeding items, leading to suboptimal performance.
 
 To address these issues, we propose a \textbf{N}on-\textbf{A}uto\textbf{R}egressive generative model for reranking \textbf{R}ecommendation (NAR4Rec) designed to enhance efficiency and effectiveness. To tackle challenges such as sparse training samples and dynamic candidates, we introduce a matching model. Considering the diverse nature of user feedback, we employ a sequence-level unlikelihood training objective to differentiate feasible sequences from unfeasible ones.  Additionally, to overcome the lack of dependency modeling in non-autoregressive models regarding target items, we introduce contrastive decoding to capture correlations among these items. Extensive offline experiments validate the superior performance of NAR4Rec over state-of-the-art reranking methods. Online A/B tests reveal that NAR4Rec significantly enhances the user experience. Furthermore, NAR4Rec has been fully deployed in a popular video app Kuaishou with over 300 million daily active users. 
 
\end{abstract}

\begin{CCSXML}
<ccs2012>
   <concept>
       <concept_id>10002951.10003317.10003338</concept_id>
       <concept_desc>Information systems~Retrieval models and ranking</concept_desc>
       <concept_significance>500</concept_significance>
       </concept>
   <concept>
       <concept_id>10010147.10010257.10010321</concept_id>
       <concept_desc>Computing methodologies~Machine learning algorithms</concept_desc>
       <concept_significance>300</concept_significance>
       </concept>
 </ccs2012>
\end{CCSXML}

\ccsdesc[500]{Information systems~Retrieval models and ranking}
\ccsdesc[300]{Computing methodologies~Machine learning algorithms}
% \ccsdesc{Theory of computation~ Online learning}

\keywords{Recommender systems, Generative Model, Non-autoregressive Models}



\maketitle

\section{Introduction}
Recommendation systems offer users personalized item lists tailored to their interests. Various approaches have been proposed to capture user interests, focusing on feature interactions\cite{cheng2016wide,guo2017deepfm,lian2018xdeepfm}, user preference modeling\cite{zhou2018deep,zhou2019deep}, and so on. However, most existing methods treat individual items separately, neglecting their mutual influence and leading to suboptimal results. Acknowledging that user interactions with one item may correlate with others in the recommendation list\cite{pei2019personalized}, reranking is introduced to consider contextual information and generate an optimal sequence of recommendation items. 

The main challenge in reranking is exploring optimal sequences within the vast space of permutations. Reranking methods are typically categorized into on-stage and two-stage approaches. One-stage methods \cite{ai2018learning,pei2019personalized} take candidates as input, estimating refined scores for each item within the permutation, and rerank them greedily based on these scores. However, one-stage methods encounter an inherent contradiction\cite{feng2021grn,xi2021context}: the reranking operation inherently alters the permutation, introducing different mutual influences between items compared to the initial arrangement. Consequently, the refined score conditioned on the initial permutation is considered implausible. 

To tackle this challenge, two-stage methods utilize a generator-evaluator framework. Here, the generator creates multiple feasible sequences, and the evaluator selects the optimal sequence based on the estimated listwise score. Within the generator-evaluator framework, the generator plays a crucial role. Generative models\cite{bello2018seq2slate,feng2021grn,gong2022real,zhuang2018globally} are preferred over heuristic methods\cite{feng2021grn,lin2023discrete,xi2021context,shi2023pier} for the generator function due to the expansive solution space of item permutations. Generative models commonly employ an autoregressive strategy for sequence generation. 

However, deploying the autoregressive models in real-time industrial recommendation systems presents challenges. Firstly, autoregressive models suffer from inference efficiency. Autoregressive models adopt a sequential approach to generate target sequences item by item, resulting in slow inference as the time complexity increases linearly with the sequence length.
Secondly, a critical issue arises from the training-inference discrepancy in autoregressive models. While these models are trained to predict the next item based on the ground truth up to that point. However, during inference, they receive their own previously generated outputs as input. This misalignment introduces an accumulated error, where inaccuracies generated in earlier timesteps propagate and accumulate over time. Consequently, this accumulation leads to sequences that deviate from the true distribution of the target sequence. Additionally, autoregressive models have limited information utilization. The sequential decoding process focuses solely on preceding items, neglecting information from succeeding items. This limitation results in suboptimal performance as the model fails to fully leverage the available context.

\begin{figure*}
    \centering
    \includegraphics[width=0.8\linewidth]{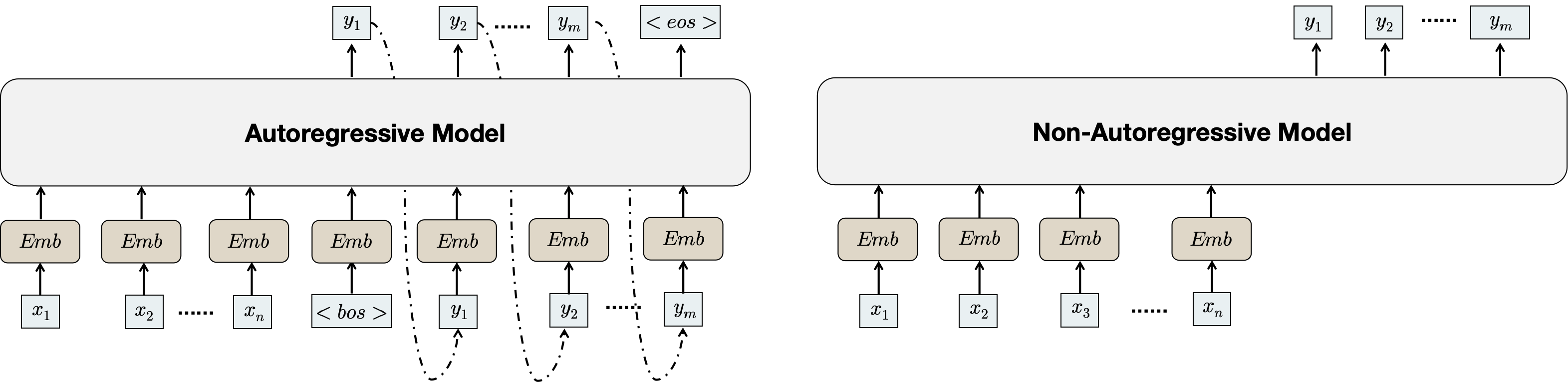}
    \caption{Comparison between autoregressive model (left) and non-autoregressive model (right). Auto-regressive models generate items sequentially while non-autoregressive models generate all items simultaneously. }
    \label{fig:compare}
\end{figure*}

To address those challenges, we propose \textbf{N}on-\textbf{A}uto\textbf{R}egressive generative model for reranking \textbf{R}ecommendation(NAR4Rec). Unlike autoregressive models, which generate sequences step by step, relying on their own previous outputs, NAR4Rec generates all items within the target sequence simultaneously. 

However, we find it nontrivial to deploy non-autoregressive in recommendation systems. Firstly, the sparse nature and dynamic candidates in recommendation systems pose difficulties for learning convergence, which we address by sharing position embedding and introducing a matching model. Secondly, the diverse nature of user feedback, including both positive and negative interactions, renders maximum likelihood training less suitable. We propose unlikelihood training to distinguish between desirable and undesirable sequences. Lastly, non-autoregressive models assume an independent selection of items at each position in a sequence, which is inadequate when modeling intra-list correlation. Hence we propose contrastive decoding to capture dependencies across items.

To summarize, our contributions are listed as follows:
\begin{itemize}
    \item We make the first attempt to adopt non-autoregressive models for reranking, which significantly speeds up the inference speed and meets the requirements of real-time recommendation systems.
    \item We propose a matching model to enhance convergence, a sequence-level unlikelihood training method to guide the generated sequence towards improved overall utility, and a contrastive decoding method to refine current decoding strategies with intra-list correlation. 
    \item  Extensive offline experiments demonstrate that NAR4Rec outperforms state-of-the-art methods. Online A/B tests further validate the effectiveness of NAR4Rec. Furthermore, NAR4Rec has been fully deployed in a real-world video app Kuaishou with over 300 million daily active users, notably improving the user experience.
\end{itemize}

\section{Related work}

\subsection{Reranking in Recommendation Systems}
In contrast to earlier phases like matching and ranking\cite{burges2005learning,liu2009learning}, which typically learn a user-specific itemwise scoring function, the core of reranking in recommendation systems lies in modeling correlations within the exposed list. Reranking\cite{pang2020setrank,pei2019personalized,xi2021context}, building upon candidate items from the ranking stage, selects a subset and determines their order to ensure exposing the most suitable items to the users. Existing research on reranking can be systematically classified into two principal categories: one-stage\cite{ai2018learning,pei2019personalized,pang2020setrank}, and two-stage methods\cite{feng2021grn,shi2023pier,lin2023discrete,xi2021context}. 

One-stage methods treat reranking as a retrieval task, recommending the top k items based on scores from a ranking model. These methods refine the initial list distribution using list-wise information, optimizing overall recommendation quality. Subsequently, the candidates are reranked by the refined itemwise score in a greedy manner. The distinction lies in the network architectures for capturing list-wise information, such as GRU in DLCM\cite{ai2018learning}, and transformer in PRM\cite{pei2019personalized}. However, user feedback for the exposed list is influenced not just by item interest but also by arrangements and surrounding context\cite{joachims2017accurately,lorigo2008eye,lorigo2006influence,yang2017relevance}. The reranking operation modifies permutations, thereby introducing influences distinct from the initial permutation. Moreover, one-stage methods, which exclusively model the initial permutation, fall short of capturing alternative permutations. Consequently, those methods struggle to maximize overall user feedback\cite{xi2021context}.

Two-stage methods \cite{feng2021grn,shi2023pier,lin2023discrete,xi2021context} embrace a generator-evaluator framework. In this framework, the generator initiates the process by generating multiple feasible sequences, and subsequently, the evaluator selects the optimal sequence based on the estimated listwise score.  This framework allows for a comprehensive exploration of various feasible sequences, and an informed selection of the most optimal one based on listwise considerations.  The role of the generator is particularly crucial for generating sequences. Common approaches for generators can be categorized into heuristic methods\cite{feng2021grn,lin2023discrete,xi2021context,shi2023pier}, such as beam search or item swapping, and generative models \cite{bello2018seq2slate,feng2021grn,gong2022real,zhuang2018globally}. Generative models are more suitable than heuristic methods for reranking given the vast permutation space. These generative models typically adopt a step-greedy strategy which autoregressively decides the display results of each position. However, the high computational complexity of online inferences limits their application in real-time recommendation systems.

To address the challenges linked with autoregressive generation models, our work investigates the viability of non-autoregressive generative models within the generator-evaluator framework.  Non-autoregressive generative models generate the target sequence once to alleviate computational complexity.

\subsection{Non-autoregressive Sequence Generation} 

Non-autoregressive sequence generation\cite{gu2018non} was initially introduced in machine translation to speed up the decoding process. Then it has since gained increased attention in nature language processing, e.g. text summarization\cite{gu2019levenshtein,awasthi2019parallel}, text error correction\cite{leng2021fastcorrect,leng2021fastcorrect2}.  Specifically, efforts have been focused on tackling the absence of target information in non-autoregressive models. Strategies include enhancing the training corpus to mitigate target-side dependencies\cite{gu2018non,zhou2019understanding} and refining training approaches\cite{ghazvininejad2019mask,stern2019insertion} to alleviate learning difficulties. 

Although non-autoregressive generation has been explored in text, those conventional techniques are not directly applicable to recommendation systems. We tackle the challenges encountered in recommendation systems to improve the convergence and performance of non-autoregressive models and make the first attempt to integrate non-autoregressive models into reranking within real-time recommender systems. 

\section{Preliminary}

\subsection{Reranking problem Formulation}
\label{sec:reranking}
For each user $u$ within the set $U$, a request encompasses a set of user profile features(such as user ID, gender, age), the recent interaction history, and $n$ candidates items denoted as $X=\left \{ x_1, x_2, \cdots, x_n \right \}$, where $n$ is the number of candidates. Given candidates $X$, the goal of reranking is to propose an item sequence $Y=\left \{ y_1, y_2, \cdots, y_m \right \}$ that elicits the most favorable feedback for user $u$, where $m$ is the sequence length and $Y$ is the recommended list of the reranking model. We denote the reranking models as $\mathcal{F}(u, \theta, X)$ where the corresponding parameter is $\theta$. In real-time recommendation systems, reranking acts as the last stage to deliver the ultimate list of recommended items. Typically, $n$ significantly exceeds $m$, with $m$ being less than 10 and $n$ ranging from several tens to hundreds.

In a multi-interaction scenario, users may exhibit distinct types of interaction (e.g., clicks, likes, comments) for each item exposure. Formally, we define the set of user interactions as $B$, and $ s_{u,y_i,b}$ represents user $u$'s response to item $y_i$ concerning interaction $b \in B$. Given ${Y}$, each item $y_i$ has a multi-interaction response $\mathbf{e}_{u,y_i} = [\textnormal{e}_{u,y_i,b_1}, \ldots, \textnormal{e}_{u,y_i,b_{|{B}|}}]$. For all items ${Y}$,the overall user response is: 
\begin{equation}
    \mathbf{E}_{u,Y} = \left[\begin{matrix}
        e_{u,y_1,b_1} & \dots & e_{u,y_m,b_1}\\  % check
        \vdots & \ddots & \vdots\\
        e_{u,y_1,b_{|B|}} & \dots & e_{u,y_m,b_{|B|}}
    \end{matrix}\right].
\end{equation}
The overall utility is quantified as the summation of individual item utilities, denoted as $\mathcal{R}(u,Y) = \sum_{i=1}^{m} \mathcal{R}(u,y_i)$. The utility associated with each item may correspond to a specific interaction type $b$, such as clicks, watch time, or likes. In such cases, the item utility is expressed as $\mathcal{R}(u,y_i) = e_{u,y_i,b}$. Alternatively, the item utility can be represented as the weighted sum of diverse interactions $\mathcal{R}(u,y_i) = \sum_{b}w_b e_{u,y_i,b}$, where $w_b$ denotes the weight for each interaction.

The reranking objective is to maximum the overall utility $\mathcal{R}(u,Y) $ for a given user $u$:
\begin{equation}
    max_{\theta} \mathcal{R}(u,Y).
\label{equ_rerank}
\end{equation}
Reranking introduces a permutation space with exponential size, represented as $\mathcal{O}(A_{n}^{m})$, where $n$ represents the number of candidates and $m$ represents the number of items to be selected and ordered. Each permutation represents a potential arrangement of items, and users provide unique feedback for each permutation. However, in practical scenarios, users typically encounter only one permutation. Thus, the main challenge in reranking lies in efficiently and effectively determining the optimal permutation given the vast solution space yet extremely sparse real user feedback as training samples.

\subsection{Autoregressive sequence generation}
Given a set of candidate items denoted as $X$, autoregressive models decompose the distribution over potential generated sequences $Y$ into a series of conditional probabilities:
\begin{equation}
    p_{\mathcal{\text{AR}}}(Y|X; \theta) = \prod_{i=1}^{m+1} p(y_i|y_{0:i-1},x_{1:n};\theta),
\label{equ_ar}
\end{equation}
where the special tokens $y_0$ (e.g., <bos>) and $y_{m+1}$ (e.g., <eos>) denote the beginning and end of target sequences. Importantly, the length of the generated sequence is predetermined and fixed, unlike variable lengths in text.

Factorizing the sequence generation output distribution autoregressively leads to a maximum likelihood training with a cross-entropy loss at each timestep:
\begin{equation}
   \mathcal{L}_{\text{AR}} = -\log p_{\text{AR}}(Y|X; \theta) = -\sum_{i=1}^{m+1}\log p(y_i|y_{0:i-1},x_{1:n};\theta).
\end{equation}
The training objective aims to optimize individual conditional probabilities. In training, when the target sequence is known, these probabilities are calculated based on earlier target items rather than model-generated ones, enabling efficient parallelization. In inference, autoregressive models generate the target sequence item-by-item  sequentially, efficiently capturing the distribution of the target sequence. This makes them well-suited for the reranking task, particularly considering the vast space of possible permutations.

While autoregressive models have proven effective, deploying them in industrial recommendation systems is challenging.  Firstly, their sequential decoding process leads to slow inference, introducing latency that hinders real-time application.  Secondly, these models, trained to predict based on ground truth, face a discrepancy during inference when they receive their own generated outputs as input. This misalignment may lead to compounded errors, as inaccuracies generated in earlier timesteps accumulate over time, resulting in inconsistent or divergent sequences that deviate from the true distribution of the target sequence. Thirdly, autoregressive models rely on a left-to-right causal attention mechanism, limiting the expressive power of hidden representations, as each item encodes information solely from preceding items\cite{sun2019bert4rec}. This constraint impedes optimal representation learning, resulting in suboptimal performance.

\subsection{Non-autoregressive sequence generation}
To address the aforementioned challenges, non-autoregressive sequence generation eliminates autoregressive dependencies from existing models. Each element's distribution $p(y_i)$ depends solely on the candidates $X$:
\begin{equation}
 p_{\text{NAR}}(Y|X; \theta) =\prod_{i=1}^m p(y_i| x_{1:n};\theta).
\label{equ_nar}
\end{equation}
Then the loss function for the non-autoregressive model is:
\begin{equation}
 \mathcal{L}_{\text{NAR}} = -\log p_{\text{NAR}}(Y|X; \theta) =-\sum_{i=1}^m \log p(y_i| x_{1:n};\theta).
\label{equ:loss_nar}
\end{equation}
Despite the removal of the autoregressive structure, the models retain an explicit likelihood function. The training of models employs separate cross-entropy losses for each output distribution.  Crucially, these distributions can be computed simultaneously during inference, which significantly differs from the sequential process of autoregressive models. This non-autoregressive approach reduces inference latency, thereby enhancing the efficiency of recommendation systems in real-world applications.

\section{Approach}
In this section, we present a detailed introduction of NAR4Rec. We will first discuss our non-autoregressive model structure, which estimates the probability by a matching model in \cref{sec:matching_model}. Then, we delve into unlikelihood training, a method aimed at discerning feedback within the recommended sequence in \cref{sec:ul_training}. Finally, we propose contrastive decoding to model the dependency in target sequence in \cref{sec:con_decoding}. The sequence evaluator in our generator-evaluator framework is introduced in \cref{sec:evaluator}.

\subsection{Matching model}
\label{sec:matching_model}
\begin{figure*}
    \centering
    \includegraphics[width=0.8\linewidth]{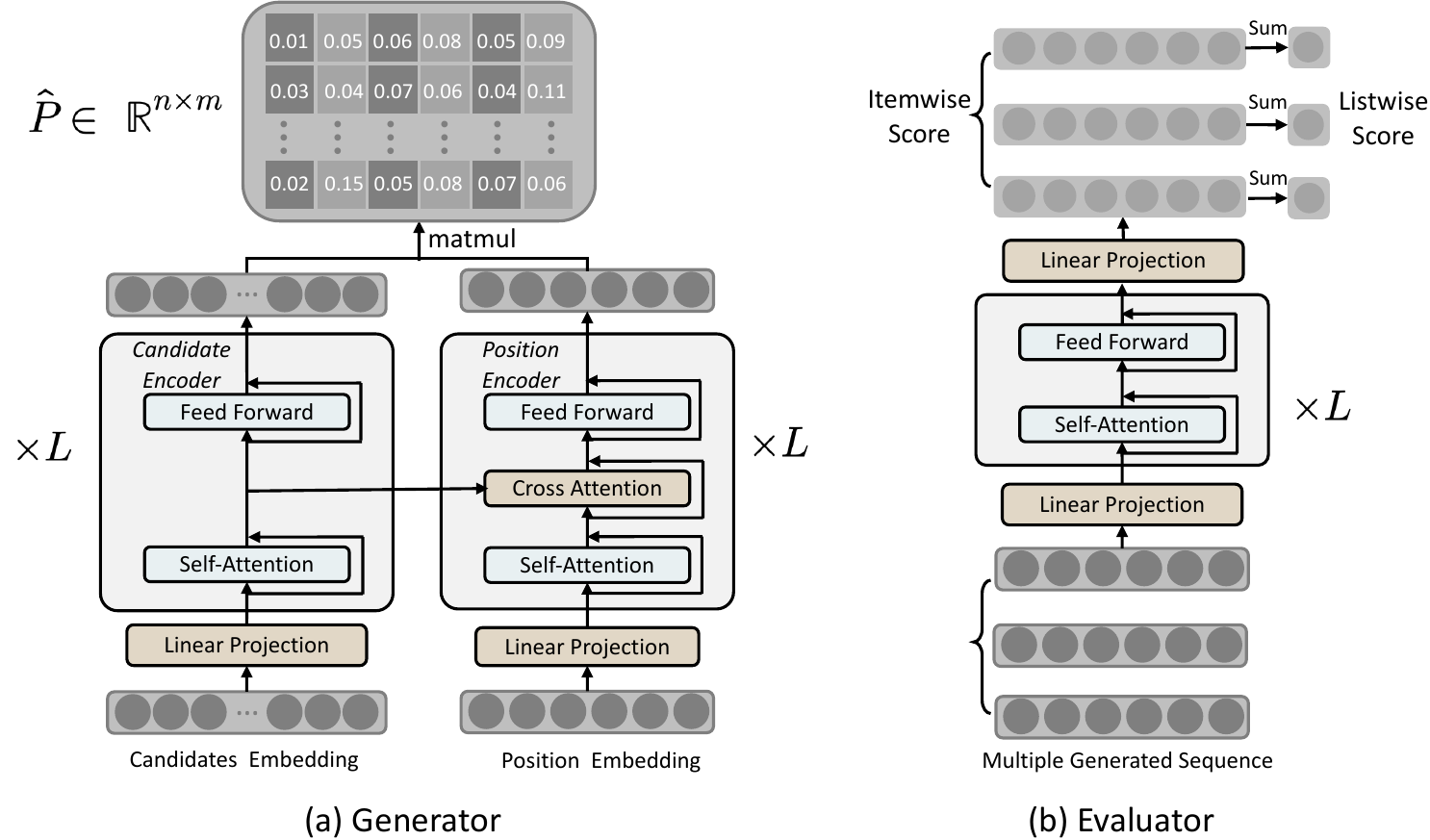}
    \caption{Architectural Overview of the Generator and Evaluator Models.  The evaluator evaluates multiple sequences generated by the generator and estimates listwise score to select the optimal sequence.}
    \label{fig:model}
\end{figure*}
Non-autoregressive models encounter challenges in training convergence due to two main reasons.  Firstly, the sparse nature of training sequences presents learning difficulties. Unlike text sequences that often share linguistic structures, recommended lists in training samples seldom have the same exposures, posing challenges for effective learning from limited data. Secondly, during the reranking stage, identical index for candidates may denote different items, leading to a variable vocabulary as the candidates to be ranked vary across samples.  

Conventional models may struggle to handle such variations efficiently. To tackle these challenges, we introduce two key components to our models: a \textit{candidates encoder} for effectively encoding representations of candidates and a \textit{position encoder} to capture position-specific information within the target sequence. Initially, we randomly initialize an embedding for each position in the target sequences. Notably, we share these position embeddings across training data to enhance learning on sparse data. Subsequently, we integrate bidirectional self-attention and cross-attention modules to acquire representations for each position, leveraging information from the candidates.  

Additionally, to address the challenge posed by the dynamic vocabulary arising from variations in candidates across samples, we employ a matching mechanism. Specifically, we match each candidate with each position in the target sequence, thereby yielding probabilities for each candidate at every position. In the following, we elaborate on the structure of NAR4Rec.

% Given a user and the candidate items as input, our model output a probability matrix $\hat{P}\in \mathbb{R}^{n\times m}$, where $\hat{P}_{i,j}$ indicates the estimated probability score of placing the $i$-th item in the $j$-th position. The model primarily consist of a candidates encoder and a position encoder. Initially, the user and item are sent to the embedding layer to be converted into item embedding. Subsequently, the candidates encoder processes these embedding by self-attention mechanism  to capture the mutual influence between items. Similarly for position encoder. Finally, the output embedding from both the candidates encoder and the position encoder are employed to compute the probability matrix through matrix multiplication.

Given a user $u$ and candidates $X=\left \{ x_1, x_2, \cdots, x_n \right \}$, the hidden representation of $x_i$ is $\mathbf{x}_i\in \mathbb{R}^{d_x}$. We stack $\mathbf{x}_i$ together into matrix $\mathbf{X}\in \mathbb{R}^{n\times d_x}$. Additionally, we randomly initialize an embedding vector for each position $j$ as $t_j\in d_{t}$. Also, we stack $t_j$ into $\mathbf{T}\in \mathbb{R}^{m\times d_t}$. To align the embedding dimension of $\mathbf{X}$ with that of  $\mathbf{T}$, we project them into same hidden dimension $d$ by a linear projection layer. Then $\mathbf{X}\in \mathbb{R}^{n\times d}$ and $\mathbf{T}\in \mathbb{R}^{m\times d}$. Consequently, $\mathbf{X}$ is represented as $\mathbf{X}\in \mathbb{R}^{n\times d}$ and $\mathbf{T}$ as $\mathbf{T}\in \mathbb{R}^{m\times d}$.

\textbf{Candidates Encoder.} The candidates encoder adopts the standard Transformer architecture\cite{vaswani2017attention} by stacking $L$ Transformer layers. In each layer, the architecture mainly consists of two blocks, a self-attention block and a feed-forward network. 
An input $\mathbf{X}$ for self-attention block is linearly transformed into query ($\mathbf{Q}_x$), key ($\mathbf{K}_x$) and value ($\mathbf{V}_x$) as follows: 
\begin{equation}
    \mathbf{Q}_x=\mathbf{X}\mathbf{W}^\mathbf{Q}_x, \mathbf{K}_x=\mathbf{X}\mathbf{W}^\mathbf{K}_x, \mathbf{V}_x=\mathbf{X}\mathbf{W}^\mathbf{V}_x, 
\end{equation}
where $\mathbf{X^Q}$, $\mathbf{X^K}$ and $\mathbf{X^V}$ denote the weight matrices. Then, the self-attention operation is applied as: 
\begin{equation}
    \text{Attention}(\mathbf{Q}_x,\mathbf{K}_x,\mathbf{V}_x)={\rm Softmax}(\frac{\mathbf{Q}_k\mathbf{K}_x^T}{\sqrt{d}})\mathbf{V}_x, 
\end{equation}

For a multi-head version of self-attention mechanism, the input is linearly projected into $\mathbf{Q}_x$, $\mathbf{K}_x$ and $\mathbf{V}_x$ with $h$ times using individual linear projections to small dimensions(e.g. $d_k=\frac{d}{h}$). Finally, the output of self-attention(SAN) is 
\begin{equation}
\begin{aligned}
    \text{SAN} &= [head_1, \cdots, head_h] \mathbf{W}^O_x, \\
    head_i &= Attention(\mathbf{Q}_i, \mathbf{K}_i, \mathbf{V}_i).
\end{aligned}
\label{equ:attention}
\end{equation}

The feed-forward network is typically placed after the self-attention block,
\begin{equation}
FFN(\mathbf{X})=\sigma(\mathbf{X}\mathbf{W}^{in}_x)\mathbf{W}^{out}_x,
\end{equation}
where $\mathbf{W}^{out}_x$ and $\mathbf{W}^{in}_x$ denote the weight matrices of the two linear projection.

\textbf{Position Encoder.} The position encoder adopts a similar Transformer architecture as the candidates encoder. The key difference between them is that the position encoder inserts a cross-attention block between the self-attention block and the feed-forward network in each Transformer layer. As can be seen in Fig.~\ref{fig:model}, in each layer, the cross-attention block receives the hidden representation from the self-attention blocks of both encoders and processes them via cross-attention operation. Specifically, the hidden representation from the candidates encoder and position encoder are denoted as $\mathbf{X}$ and $\mathbf{T}$, respectively. Similar to the self-attention block, we initially apply linear projection to them: 
\begin{equation}
    \mathbf{Q}=\mathbf{T}\mathbf{W}^\mathbf{Q}, \mathbf{K}=\mathbf{X}\mathbf{W}^\mathbf{K}, \mathbf{V}=\mathbf{X}\mathbf{W}^\mathbf{V}. 
\end{equation}
Then, we applies the formula in \cref{equ:attention} to $\mathbf{Q}$, $\mathbf{K}$, and $\mathbf{V}$ to get the output hidden representation. The cross-attention is introduced to capture the correlation between candidates and target sequence.

\textbf{Probability Matrix.} To compute the probability matrix, we perform a matrix multiplication on the output hidden representation from the candidates encoder (denoted as $\{\mathbf{x_1}, \mathbf{x_2}, ..., \mathbf{x_n}\}$) and position encoder (denoted as $\{\mathbf{t_1}, \mathbf{t_2}, ..., \mathbf{t_m}\}$). Subsequently, we apply a column-wise softmax function to normalize the scores. Formally, the probability score of placing the $i$-th candidate item to the $j$-th position is calculated as: 
\begin{equation}
    \hat{p_{ij}} = \frac{exp(\mathbf{x_i}^\intercal \mathbf{t_j})}{\sum_{i=1}^{n}exp(\mathbf{x_i}^\intercal \mathbf{t_j})}.
\end{equation}

\textbf{Training objectives} NAR4Rec is trained via cross-entropy loss function, defined as follows:
\begin{equation}
    \mathcal{L}(Y,X) = -\sum_{i=1}^{n}\sum_{j=1}^{m}  p_{ij} \log (\hat{p_{ij}}),
\end{equation}
where $p_{ij}$ is 1 if $x_j$ is in position $p_j$ otherwise 0.

\subsection{Unlikelihood Training}
\label{sec:ul_training}
The discrepancy between text and item sequences hinders the direct application of generative models from text to item recommendation. This disparity arises from the unique characteristics of user interactions in recommendation scenarios.  Unlike the structured nature of natural language text, the feedback within recommendation sequences is diverse due to the varied nature of user interactions. While text sequences typically follow conventional language structures to convey information or construct a coherent narrative, user feedback in recommendation sequences is characterized by diverse actions such as clicks or likes,  reflecting a diverse and nuanced feedback.

Consequently, the difference in objectives between maximum likelihood training(as in \cref{equ_nar}) and reranking (as in \cref{equ_rerank}) poses a significant challenge. Although maximum likelihood training effectively captures patterns in text sequences, its applicability diminishes in recommendation scenarios where user preferences are dynamic and subjective. The essence of high-quality recommendations lies not just in sequence patterns from training data but, more crucially, in the user utility of the recommended list. User interactions with recommended items are subjective and context-driven, adding complexity to aligning the training objective with the desired model behavior. To address this challenge, we propose unlikelihood training, guiding the model to assign lower probabilities to undesired generations. This adjustment aligns the training process with the intricate feedback patterns.

Unlikelihood training reduces the model's probability of generating a negative sequence. Given candidates $X$ and a negative sequence $Y_{neg}$, the unlikelihood loss is:
\begin{equation}
    \mathcal{L}_{\text{ul}}(Y_{neg},X) = -\sum_{i=1}^{n}\sum_{j=1}^{m} p_{ij} \log (1-\hat{p}_{ij}),
\end{equation}
The loss decreases as $\hat{p}_{ij}$ decreases.

Unlike text generation, where messages are clear and content-focused, managing attributes like topic, style, and sentiment in the output text is straightforward\cite{li2020don,welleck2019neural}. However, recommendation sequences involve user feedback with implicit signals. For instance, a lack of interaction with a recommended item may suggest disinterest. This highlights the model's need to understand both explicit and implicit cues in user feedback.  Effective control over generation in recommendation sequences becomes crucial to tailor the output based on user preferences and behaviors, thus enhancing the personalized recommendations. Specifically, we classify a item sequence as positive or negative based on the overall utility defined in \cref{sec:reranking}, and the corresponding loss is as follows:
\begin{equation}
    \mathcal{L}_{\text{ul}}(Y,X)=\left\{\begin{array}{ll} -\sum_{i=1}^{n}\sum_{j=1}^{m} p_{ij} \log (1-\hat{p}_{ij}) & \text { if } \mathcal{R}(u,Y)<\tau \\ -\sum_{i=1}^{n}\sum_{j=1}^{m}  P_{ij}\log (\hat{P_{ij}})  & \text { if } \mathcal{R}(u,Y)\geq \tau\end{array}\right.
\label{equ:ul}
\end{equation}
where $\alpha$ is the threshold for positive and negative sequences. 

In summary, beyond the primary goal of learning positive sequence patterns through sequence likelihood, unlikelihood training introduces an additional objective to reduce the likelihood of generating sequences with low utilities, effectively training recommendation models to discern feedback within recommendation sequences.

\subsection{Contrastive Decoding}
\label{sec:con_decoding}
Compared with autoregressive generation, the non-autoregressive approach significantly enhances computation efficiency and makes it feasible to deploy in real-time recommendation systems. However, non-autoregressive generation introduce the conditional independence assumption: each target item's distribution $p(y_i)$ depends only on the candidates $X$. This deviation from autoregressive models poses challenges in capturing the inherently multimodal nature of the distribution of valid target sequences. Take machine translation for example, when translating the phrase "thank you" into German could result in multiple valid translations such as "Vielen Dank" and "Danke". However, non-autoregressive models may generate unplausible translations like "Danke Dank". The conditional independence assumption in \cref{equ_nar} restricts the model's ability to effectively grasp the multimodal distribution in target sequences.

Essentially, the assumption of conditional independence limits the model's ability to navigate a vast solution space and identify the most suitable permutation from numerous valid options for a given set of candidates. This limitation is especially evident in recommendation where the number of reasonable target sequences far exceeds those encountered in text. Consequently, non-autoregressive frameworks grapple with  the challenge of mitigating the impact of conditional independence to improve their capacity for generating diverse and contextually appropriate target sequences. To tackle this, we propose contrastive decoding to model the co-occurrence relationship between items and thereby improve the target dependency. 

Contrastive decoding incorporates a diversity prior that regulates the sequence decoding procedure. This is grounded in the intuition that an effective recommended list needs to be composed of a wide variety of items. In fact, contrastive decoding leverages a similarity score function as a regulator when decoding, capturing the interdependence between various positions in the target sequence.

% Contrastive decoding introduces a transition matrix to capture dependencies between different positions in the target sequence. By incorporating this matrix into the generation process, non-autoregressive models can account for dependencies between positions, improving the coherence of the generated sequences.The rationale behind introducing the transition matrix is to enhance the model's understanding of relationships within the target sequence. This approach not only reduces the impact of conditional independence but also enables the model to generate sequences that are more informed and contextually aware.

Formally, given the preceding context $y_{<i}$, at time step $i$, the selection of the output $y_i$ follows
\begin{equation}
    y_t = argmax_{x \in X} (1-\alpha)\times p(x|p_i,X)-\alpha \times max(s(\mathbf{x},\mathbf{x}_j)),
\label{equ:contrastive_decoding}
\end{equation}
where $0 \leq j \leq t-1$. In \cref{equ:contrastive_decoding}, the first term, termed as model confidence, denotes the probability of candidates $x$ predicted by the model. The second term, known as similarity penalty, quantifies  the distinctiveness candidate $x$ concerning the previously selected items, where $s(\mathbf{x}, \mathbf{x_j})$ is computed as:
\begin{equation}
    s(\mathbf{x}, \mathbf{x_j}) = \frac{\mathbf{x}^\top \mathbf{x_j}}{\|\mathbf{x}\|\cdot\|\mathbf{x_j}\|}.
\end{equation}
Specifically, the similarity penalty is defined as the maximum similarity between the representation of $x$ and all items in $y_{<i}$. NAR4Rec utilizes the dot product item embedding and position embedding to compute the probability matrix. Higher embedding similarity between items often means similar probability in a certain position. We introduce such penalty to introduce intra-list correlation.

Also, to encourage the language model to learn discriminative and isotropic item representations, we incorporate a contrastive objective into the training of the language model. Specifically, given a sequence $X$, the $\mathcal{L}_{\text{item}}$ and $\mathcal{L}_{\text{position}}$ are defined as:
\begin{equation}
    \label{eq:cl}
    \mathcal{L}_{\text{item}} = \frac{1}{n\times(n - 1)}\sum_{i=1}^{n}\sum_{j=1,j\neq i}^{n}\max\{0,\rho - s(\mathbf{x}_i, \mathbf{x}_i) + s(\mathbf{x}_i, \mathbf{x}_j)\},
\end{equation}
where $\rho\in[-1,1]$ is a pre-defined margin  and $x_i$ is the hidden representation of item $x_i$ from candidates encoder. 
\begin{equation}
    \label{eq:pos_cl}
    \mathcal{L}_{\text{position}} = \frac{1}{m\times(m - 1)}\sum_{i=1}^{m}\sum_{j=1,j\neq i}^{|\boldsymbol{x}|}\max\{0,\rho - s(\mathbf{t}_i, \mathbf{t}_i) + s(\mathbf{t}_i, \mathbf{t}_j)\},
\end{equation}
where $t_i$ is the hidden representation of position $t_i$ from the position encoder.
Intuitively, by training with $\mathcal{L}_{\textup{CL}}$, the model learns to pull away the distances between representations of distinct tokens.\footnote{By definition, the cosine similarity $s(h_{x_i}, h_{x_i})$ of the identical token $x_i$ is $1.0$.} Therefore, a discriminative and isotropic model representation space can be obtained. 

The overall training objective $\mathcal{L}$ is then defined as
\begin{equation}
    \label{eq:simctg}
    \mathcal{L}(Y,X) = \mathcal{L}_{\text{ul}} + \mathcal{L}_{\text{position}}+\mathcal{L}_{\text{item}},
\end{equation}
where the unlikelihood training objective is described in \cref{equ:ul}. Note that, when the margin $\rho$ in $\mathcal{L}_{\text{position}}$ and $\mathcal{L}_{\text{item}}$  equals to $0$, the $\mathcal{L}(Y,X)$ degenerates to the vanilla unlikelihood training objective $\mathcal{L}_{\textup{ul}}$.

\subsection{Sequence Evaluator}
\label{sec:evaluator}
The sequence evaluator model is designed to estimate the overall utility of a given sequence, as illustrated in \cref{fig:model}. The generated sequence from the generator is first encoded using a self-attention and a feed-forward layer to capture contextual information. The hidden representation then passes through the linear projection layer to predict the score for a specific target. The overall utility is calculated as the weighted sum of itemwise scores. Ultimately, the sequence with the highest overall utility is chosen for delivery to the users.

\section{Experiments}
In this section, we conduct extensive offline experiments and online A/B tests to demonstrate the effectiveness of NAR4Rec. 
We first describe our experiment setup and baselines in  \cref{sec:experiment_over}. For offline experiments in \cref{sec:offline}, we  compare NAR4Rec with existing baselines on both performance and training and inference time. Then we alternate the hyper-parameter to analyse the hyper-parameter sensitivity of NAR4Rec. To further show the effectiveness of NAR4Rec in real-time recommendation system, we conduct online A/B tests to ablate our proposed methods in \cref{sec:online}.

% We conduct extensive experiments in both offline and online environments to demonstrate the effectiveness of NAR4Rec. Three questions are investigated in the experiments:
% \begin{itemize}
%     \item How does NAR4Rec perform in comparison with state-of-the-art performance in terms of recommendation quality?
%     \item How do different settings of NAR4Rec affect the performance?
%     \item How does NAR4Rec perform in real-life recommendation systems?
% \end{itemize}

\subsection{Experiment details}
\label{sec:experiment_over}
\begin{table}[ht]
  \caption{Statistics of the datasets. \ \ }
  \renewcommand\arraystretch{1.1}
  \centering
  \setlength{\tabcolsep}{3.5mm}{
  \begin{tabular}{c|ccc}
    \hline
  Dataset & \#Requests  &  \#Users  & \#Ads \\
  \hline
  \hline
  Avito & 53,562,269  &1,324,103   &  23,562,269 \\
  Meituan & 230,525,531  &3,201,922  & 98,525,531 \\
  \hline
  \end{tabular}
  }
  \label{table:statistics}
\end{table}

\textbf{Dataset}: To evaluate reranking recommendation, we expect that each sample of the dataset is an exposed sequence to users rather than a manually constructed sequence. For public dataset, we choose Avito dataset. For industrial dataset, we use real-world data collected from Kuaishou short-video platform. The detailed introduction is given in \cref{table:statistics}.

\begin{itemize}
\item \textbf{Avito}\footnote{https://www.kaggle.com/c/avito-context-ad-clicks/data}:  The Avito dataset is a publicly available collection of user search logs from avito.ru. The dataset comprises over 53 million lists with 1.3 million users and 36 million ads. Each sample corresponds to a search page with multiple ads. The user search logs from first 21 days are used as training set and the search logs from the last 7 days are used as test set. The sequence length in Avito is 5.
    
\item \textbf{Kuaishou}: The Kuaishou dataset is derived from Kuaishou, a widely used short-video application with a user base of over 300 million daily active users. Each sample in the dataset represents an actual user request log, which contains user information(e.g. user id, age, gender), candidates items and user interaction to exposed items. The dataset consists of a total of 82,230,788 users, 26,835,337 items, and 1,811,625,438 requests. Each request contains 6 items in the exposed item sequence and 60 candidates from ranking.
\end{itemize}

\subsection{Offline experiments}
\label{sec:offline}
\textbf{Baselines} We compare the proposed NAR4Rec with 6 state-of-the-art reranking methods. We select DNN, DCN as pointwise baselines, PRM as one-stage listwise baselines, Edge-Rerank, PIER, Seq2slate as two-stage baselines. Crucially, Seq2slate is a autoregressive generative model. A brief overview of these baseline methods is as follows:
\begin{itemize}
    \item \textbf{DNN}\cite{covington2016deep}: DNN is a basic method for click-through rate prediction, which applies a multi-layer perception to learn feature interaction.
    \item \textbf{DCN}\cite{wang2017deep}: DCN incorporates feature crossing at each layer, eliminating the need for manual feature engineering while keeping the added complexity to the DNN model minimal.
    \item \textbf{PRM}\cite{pei2019personalized}: PRM models the mutual correlation among items by leveraging the self-attention mechanism and then rank the items by the estimated scores to generate the item sequence.
    \item \textbf{Edge-Rerank}\cite{gong2022real}: Edge-Rerank generates the context-aware sequence with adaptive beam search on estimate scores.
    
    \item \textbf{Seq2Slate}\cite{bello2018seq2slate}: Seq2Slate leverages pointer networks, which are seq2seq models with an attention mechanism to predict the next item given the items already selected.
    % Seq2Slate leverages pointer networks to capture item dependencies while maintaining computational efficiency.
     \item \textbf{PIER}\cite{shi2023pier}: PIER applies hashing algorithm to slect top-k candidates from the full permutation based on user interests.  Then the generator and evaluator are jointly trained to generate better permutations.
\end{itemize}

% \textbf{Implementation Details} For different datasets, we use different user feedback as condition. For Avito, blabla

\textbf{Metrics} As there is not common sequence generation metrics for recommendation, we follow previous work\cite{shi2023pier,lin2023discrete} and evaluate these models using three commonly adopted metrics: AUC, LogLoss, and NDCG on Avito dataset. For Avito dataset, where $n$ and $m$ are equal(5), the task is to predict the itemwise click-through rate given a listwise input. For Kuaishou dataset, where $n$ and $m$ are $60$ and $6$ respectively, we employ Recall@6, Recall@10, and LogLoss as evaluation metrics. The task for Kuaishou dataset is to predict whether an item is chosen to be one of the exposed $6$ items.

\textbf{Hyperparameters} For Avito dataset, the learning rate is 10-3, the optimizer is Adam and the batch size is 1024. For Kuaishou dataset, the learning rate and optimizer is the same as Avito, but the batch size is 256. 

% To do list
% \begin{itemize}
%     \item Hyper-parameter experiments (learning rate, epochs, batch size) [done]
%     \item Comparison with Seq2slate(important) [done], Set2rank, DLCM, EGRerank(Edge-reranking) [done]
%     \item Metrics(NDCG) [done],NAR4Rec(alpha) [done]
%     \item Computation complexity(Seq2slate vs NAR4Rec) [done]
%     \item Decoding methods, Unlikelihood loss[done]
%     \item Writing
% \end{itemize}

\subsubsection{Performance comparison} Here we show the results of our proposed method NAR4Rec. As can be seen in in \cref{tab:avito_dataset} and \cref{tab:industry_dataset},  NAR4Rec outperforms 5 baslines including recent strong reranking methods\cite{pei2019personalized, bello2018seq2slate, shi2023pier}. PRM outperforms DNN and DCN by effectively capturing the mutual influence between items. Additionally, Edge-rerank surpass DNN and DCN with an adative beam search with previous item information. PIER demonstrates superiority over other baselines by the interaction per category. Notably, our proposed method exhibits the highest improvement with a significant increase of 0.0125 in the AUC metric compared to other baseline models.  \cref{tab:industry_dataset} shows the results of our offline experiments on Kuaishou. The evaluation metrics used in this experiment include Recall@6, Recall@10, and Loss. Our method achieves superior results on all metrics compared to other baseline models as well.

\begin{table}[ht]
    \centering
    \caption{Comparison between Nar4Rec and baseline methods on the Avito dataset.}
    \begin{tabular}{c|cccc}
    \toprule
            & AUC &  LogLoss & NDCG\\
    \midrule
        DNN &  0.6614 & 0.0598 & 0.6920\\
        DCN &  0.6623 & 0.0598 & 0.7004\\
        PRM &  0.6881 & 0.0594 & 0.7380\\
        Edge-rerank &  0.6953 & 0.0574 & 0.7203\\
        PIER &  0.7109 & 0.0409 & 0.7401 \\
        Seq2Slate & 0.7034 & 0.0486 & 0.7225\\
        % Setrank & & \\
        NAR4Rec & \textbf{0.7234} & \textbf{0.0384} & \textbf{0.7409} \\
    \bottomrule
    \end{tabular}
    \label{tab:avito_dataset}
\end{table}

\begin{table}[ht]
    \centering
    \caption{Comparison between Nar4Rec and baseline methods on KuaiShou.}
    \begin{tabular}{c|cccc}
    \toprule
        & Recall@6 & Recall@10 & LogLoss \\
    \midrule
        DNN & 66.47\% & 86.65\% & 0.6764 \\
        DCN & 68.22\% & 87.95\% & 0.6809 \\
        PRM & 73.17\% & 92.25\% & 0.5328 \\
        Edge-rerank & 73.63 \%& 92.90\%& 0.5252 \\
        PIER & 73.50\% & 92.44\% & 0.5361 \\
        NAR4Rec & \textbf{74.86\%} & \textbf{93.16\%} &  \textbf{0.5199} \\
    \bottomrule
    \end{tabular}
    \label{tab:industry_dataset}
\end{table}

\subsubsection{Training and Inference Time comparison}

Given that NAR4Rec is closely related to autoregressive models, we conduct a comparison with autoregressive models Seq2Slate. We compare the training and inference time on the Avito dataset between Seq2slate and NAR4Rec. We also give training and inference time for generators in other baselines in  \cref{tab:training_time}.   Since Seq2slate utilizes recurrent neural networks as its backbone network, both training and inference processes adopt an autoregressive manner. The inference speedup of NAR4Rec over Seq2slate is almost the same as training. NAR4Rec only requires 58 minutes to complete the training while Seq2Slate requires 283 minutes. Such a significant reduction in training time (i.e. approximately 5$\times$ speedup) highlights the computational efficiency of NAR4Rec. The autoregressive model represented by Seq2Slate generates target sequences item by item, while our Non-autoregressive NAR4Rec generates all items at once. So when generating a sequence of length 5, NAR4Rec shows approximately 5$\times$ speedup.

\begin{table}[ht]
    \centering
    \caption{The training and inference time comparison between NAR4Rec and  baseline methods on the Avito dataset. All experiments are conducted on Tesla T4 16G GPU and the batch size is set to 1024. The training and inference time is calculated by averaging the result over 100 steps.}
    \begin{tabular}{c|cc}
    \toprule
         & Training Time & Inference Time \\
    \midrule
    DNN & \textbf{0.102s} & \textbf{0.034s} \\
    DCN & 0.106s & 0.035s  \\
    PRM & 0.109s & 0.036s  \\
    Edge-rerank & 0.105s & 0.035s   \\
    PIER & 0.160s & 0.053s \\ %only generator
    Seq2Slate & 0.558s & 0.186s  \\
    NAR4Rec & 0.112s & 0.037s  \\ %only generator
    \bottomrule
    \end{tabular}
    \label{tab:training_time}
\end{table}

\subsubsection{Hyper-parameter Analysis of NAR4Rec}
\begin{figure*}
    \centering
    \includegraphics[width=0.8\linewidth]{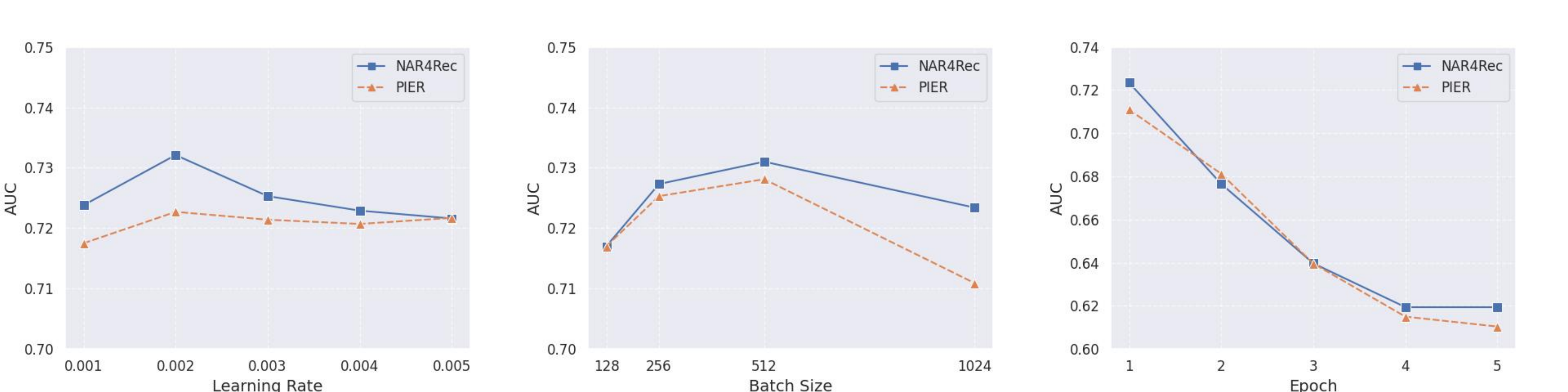}
    \caption{The comparison between NAR4Rec and PIER on Avito with different learning rate, batch size and epoch.}
    \label{fig:combine}
\end{figure*}

We further analyze the hyper-parameter sensitivity on NAR4Rec. Here, we conduct a series of experiments on NAR4Rec and PIER. As shown in \cref{fig:combine}, we demonstrate that our experimental results exhibit insensitivity to variations in the learning rate, batch size, and epoch.

\begin{figure*}[ht]
    \centering
    \includegraphics[width=0.8\linewidth]{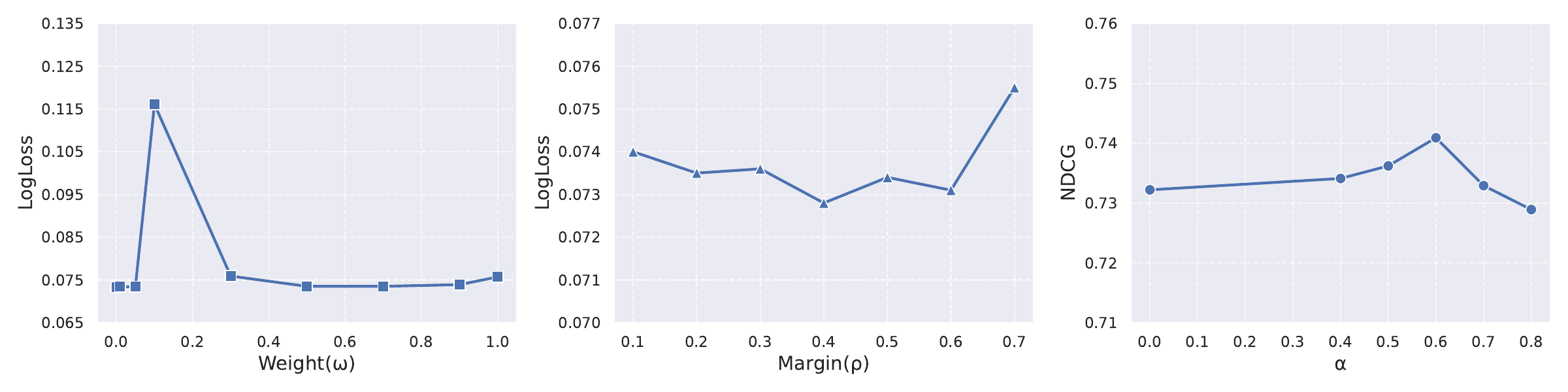}
    \caption{The effect of parameter weight $\omega$ and margin $\rho$ in contrastive loss and the $\alpha$ in contrastive decoding.}
    \label{fig:wei_rho_alpha}
\end{figure*}

Then, we analyze the impact of weight $\alpha$ and margin $\rho$ in contrastive loss and the impact of penalty parameter $\alpha$ in contrastive decoding. \Cref{fig:wei_rho_alpha} shows the results of our experiments. We change $\omega$ while fixing $\rho$=0.5, and change $\rho$ while fixing $\omega$=0.01 in contrastive loss. When changing $\alpha$ in contrastive decoding, we set $\rho$=0.5 and $\omega$=0.01 as the default parameters. 

\subsection{Online experiments}
\label{sec:online}
Text sequence generation often is evaluated by human labeling. In recommendation, we resort to online A/B experiments to obtain the feedback from users to demonstrate our effectiveness. 

\subsubsection{Experiments setup}
In online A/B experiments, we evenly divide the traffic of the entire app into ten buckets. The online baseline is Edge-rerank\cite{gong2022real}, with 20\% of the traffic assigned to NAR4Rec, while the remaining traffic is assigned to Edge-rerank. 

\subsubsection{Experimental Results}
The experiments have been launched on the system for ten days, and the result is listed in \cref{tab:online}. NAR4Rec outperforms Edge-rerank\cite{gong2022real} by a large margin. NAR4Rec shows users watch more(i.e a higher Views) videos, spend more time on each video(i.e. more Long Views and Complete Views) and a more positive user feedback(i.e. a improvement on like, follows) over  \cite{gong2022real}. 

\begin{table}[ht]
\begin{tabular}{ccccc}
\toprule
Views & Likes & Follows  & Long Views & Complete Views \\
\midrule
    +1.161\% & +1.71\%  & +1.15\% & +1.82\% &  +2.45\%\\
\bottomrule
\end{tabular}
\caption{ Online experiments results. All values are the relative improvements of NAR4Rec. For the online A/B test in Kuaishou, the improvements of over 0.5\% in positive interactions(like, follow) and 0.2\% in views are very significant.}
\label{tab:online}
\end{table}

\subsubsection{Ablation Study on Unlikelihood Training}
To show the effectiveness of unlikelihood training, we compare vanilla training on all exposed sequences with unlikelihood training. Unlikelihood shows more Views and a longer Watch Time.
\begin{table}[ht]
\begin{tabular}{c|cc}
\toprule
    & Views & Watch Time  \\
\midrule
 Vanilla training   & -0.370\%* & -0.277\%* \\
\bottomrule
\end{tabular}
\caption{Online experiments on training methods. All values are the relative changes over Unlikelihood Training. * indicates  that the metrics is statistically significant.}
\label{tab:training_methods}
\end{table}
\subsubsection{Ablation Study on Contrastive Decoding}
Here we compare the common decoding algorithm in text sequence and a diversity-based algorithm(i.e. Deep DPP)with contrastive decoding. Those decoding algorithms shows a significant drop in View and Watch time, suggesting  a poorer user feedback.

\begin{table}[ht]
\begin{tabular}{c|cc}
\toprule
    & Views & Watch Time  \\
\midrule
 Deep DPP   & -0.363\%* & -0.361\%* \\
 Beam Search & -0.327\%* &  -0.214\%* \\
 Greedy Search & -0.216\%* &  -0.178\%* \\
 Top-k Sampling & -0.254\%* & -0.131\% \\
\bottomrule
\end{tabular}
\caption{Online experiments on decoding methods. All values are the relative changes over the Contrastive Decoding. * indicates  that the metrics is statistically significant.}
\label{tab:decoding_methods}
\end{table}

\section{Conclusion}
In this paper, we provide an overview of the current formulation and challenges associated with reranking in recommendation systems. Although non-autoregressive generation has been explored in natural language processing, conventional techniques are not directly applicable to recommendation systems. We tackle the challenges in recommendations to improve the convergence and performance of non-autoregressive models and make the first attempt to integrate non-autoregressive models into reranking in real-time recommender systems. Extensive online and offline A/B experiments have demonstrated the effectiveness and efficiency of NAR4Rec as a versatile framework for generating sequences with enhanced utility. Moving forward, our future work will focus on refining the modeling of sequence utility to further enhance the capabilities of NAR4Rec.

\bibliographystyle{ACM-Reference-Format}
\balance
\bibliography{sample-base}

\appendix

\end{document}